\documentclass[9pt,conference]{IEEEtran}

\usepackage[utf8]{inputenc}
\usepackage{graphicx} % figures
\usepackage{float} % correct placement of figures
\usepackage{subfig}
\usepackage{todonotes} % development
\usepackage{wrapfig} % text umfluss
\usepackage{rotating} % rotation von grafiken
\usepackage{verbatim} % multiline comments
\usepackage{mathtools} % equations
\usepackage{latexsym} % additional symbols
\usepackage{listings} % quelltext
\usepackage[linesnumbered]{algorithm2e} % pseudocode
\usepackage{amsfonts} % mathbb environment
\usepackage{pifont}
\usepackage[square,sort,comma,numbers]{natbib} % quellen sortieren, ranges
\usepackage{amsmath} % split environment

\lstdefinestyle{mystyle}{  
    commentstyle=\color{gray},
    keywordstyle=\color{blue},
    stringstyle=\color{purple},
    basicstyle=\ttfamily\tiny,
    breakatwhitespace=false,         
    breaklines=false,                 
    captionpos=b,                    
    keepspaces=true,                 
    showspaces=false,                
    showstringspaces=false,
    showtabs=false,                  
    tabsize=2
}

\usepackage{tikz}
\newcommand*\circled[1]{\tikz[baseline=(char.base)]{
            \node[shape=circle,draw,inner sep=2pt] (char) {#1};}}
            
\usepackage{enumitem} % Beschreibungen anpassen
\usepackage{textcomp}
\usepackage{balance}
\usepackage{eso-pic} % accepted-manuscript notice on the first page
\usepackage[hidelinks]{hyperref} % DOI link in the accepted-manuscript notice

\newcommand{\IEEEAcceptedManuscriptNotice}{%
  \AddToShipoutPictureFG*{%
    \AtPageLowerLeft{%
      \put(48,9){%
        \parbox[b]{516pt}{%
          \hrule
          \vspace{2pt}
          \fontsize{5.2}{5.8}\selectfont
          \raggedright
          \textcopyright~2019 IEEE. Personal use of this material is permitted.
          Permission from IEEE must be obtained for all other uses, in any current
          or future media, including reprinting/republishing this material for
          advertising or promotional purposes, creating new collective works, for
          resale or redistribution to servers or lists, or reuse of any copyrighted
          component of this work in other works.

          \textit{This is the accepted manuscript of: J. Novacek, A. Viehl,
          O. Bringmann, and W. Rosenstiel, ``Ontology-based Requirements
          Transformation,'' in 2019 IEEE International Symposium on Systems
          Engineering (ISSE), pp. 1--8, 2019. The version of record is available at
          \href{https://doi.org/10.1109/ISSE46696.2019.8984265}{%
          \textcolor{blue}{https://doi.org/10.1109/ISSE46696.2019.8984265}}.}%
        }%
      }%
    }%
  }%
}
\begin{document}
\IEEEAcceptedManuscriptNotice

\title{Ontology-based Requirements Transformation}
%

%\author{\IEEEauthorblockN{- Omitted for blind review -
%\vspace{1.85cm}}}

\author{\IEEEauthorblockN{Jan Novacek\IEEEauthorrefmark{1}\IEEEauthorrefmark{2}, Alexander Viehl\IEEEauthorrefmark{1}, Oliver Bringmann\IEEEauthorrefmark{1}\IEEEauthorrefmark{2}, Wolfgang Rosenstiel\IEEEauthorrefmark{1}\IEEEauthorrefmark{2}}
\IEEEauthorblockA{\IEEEauthorrefmark{1}FZI Forschungszentrum Informatik\\
Haid-und-Neu-Str. 10-14, 76131 Karlsruhe}
\IEEEauthorblockA{\IEEEauthorrefmark{2}Eberhard Karls Universität Tübingen\\
Sand 14, 72076 Tübingen}}

\begin{comment}

\author{Jan Novacek\inst{1,2} \and Alexander Viehl\inst{2}}
%
%\authorrunning{Jan Novacek et al.} % abbreviated author list (for running head)
%\tocauthor{Jan Novacek, Alexander Viehl}
%
\institute{University of Tübingen,\\
Geschwister-Scholl-Platz, 72074 Tübingen, Germany\\
\email{novacek@fzi.de},\\
\texttt{http://www.fzi.de/mitarbeiter/novacek}
\and
FZI Research Center for Information Technology,\\
Haid-und-Neu-Str. 10-14, 76131 Karlsruhe, Germany}

\end{comment}

\maketitle              % typeset the title of the contribution

\begin{abstract}
This paper presents an ontology-based approach to the supply chain-aware transformation of functional and environmental load requirements given by so-called Mission Profiles (MPs). The approach aims at improving the efficiency of the engineering process through supporting the transformation process and enabling a better integration of the transformation into existing Model-based Systems Engineering (MBSE) processes. We propose a methodology and a supporting system which aids in the transformation process while the latter feature is obtained by constructing and working on models. Consequent utilization of the standardized language OWL to express model representations further enables better knowledge integration and transfer among heterogeneous systems. In addition to that, this favors knowledge reuse across projects which can reduce overall costs. Moreover, the system enables stripping off irrelevant information from MPs, thus improving protection of intellectual property.
\end{abstract}

\section{Introduction}
\label{sec:introduction}
% Intro
One important aspect of the tremendous increase of system complexity is the resulting necessity to manage huge amounts of requirements efficiently. This includes the processing of requirements for transforming them as they are broken down to separate tier needs. The data transformation framework presented in this paper aims specifically at improving the efficiency of the engineering process through supporting the process of transforming functional and environmental load requirements given by so-called Mission Profiles (MPs). Ontologies are used to capture knowledge and to represent models to enable a better integration of heterogeneous data sources and to encourage reuse of knowledge across project life cycle phases and between projects which has been identified as an important aspect of today's Systems Engineering challenges \citep{ambrosio2017mbseChallenges}.

\begin{figure}
  \centering
  \includegraphics[width=\linewidth]{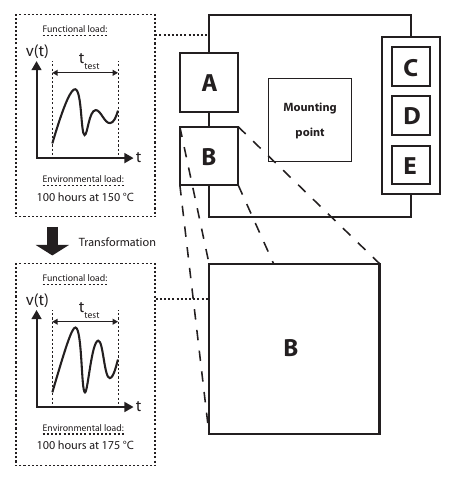}
  \captionsetup{width=\linewidth}
  \caption[bla]{MP transformation is necessary due to the difference in functional and environmental loads of port B of a component with a single mounting point}
  \label{fig:mission_profile_transformation}
\end{figure}

% Approach specific descriptions
Typically, requirements are passed along the supply chain while being refined and broken down to subsequent tier needs. In this process MP data needs to be transformed, see Fig. \ref{fig:mission_profile_transformation}. The selection of an appropriate transformation, its parametrization, and applying it to the MP data can be supported by using ontologies and reasoning. Using MBSE methodologies is accepted as an effective way in the design process of complex systems in the automotive and aerospace industries \citep{ambrosio2017mbseChallenges,wang2016effort}. The presented approach follows the idea of using and expressing models through the standardized ontology language OWL \cite{w3dc2012owl} and the Semantic Web Rule Language (SWRL) \cite{oconnor2005swrlapi} to further improve knowledge sharing and reuse among heterogeneous systems. Therefore, the approach can be integrated into existing MBSE methods. It supports multiple Model-to-Model (M2M) transformation approaches and is extensible.
% General descriptions
Basically, this paper aims at providing support for the process of transforming MPs. We investigate on whether and if so how Semantic Web Technologies (SWTs) can be used to accomplish this, focusing on the use of ontologies.

% Problems
The design of complex systems imposes several engineering challenges. A predominant problem is that Systems Engineering data is primarily stored in documents that are scattered across engineering workstations, and poorly managed \citep{ambrosio2017mbseChallenges}. With regard to requirements, MPs are shared in a non-standardized way for example by using Excel sheets or Word documents. Transforming them automatically is therefore challenging as it requires implementing a custom solution for each MP representation.
Another problem is the consideration of system information and the integration of relevant data into the transformation process.

% Solutions
This paper addresses the aforementioned problems through the proposal of a system which is capable of integrating heterogeneous data sources to build up models. The main goal of supporting MP transformation is then carried out using the constructed models.
The benefits are primarily:
\begin{itemize}
\item Better protection of intellectual property (IP) due to removal of non necessary information
\item Decreased communication overhead due to less transformations
\item Better sharing and reuse of transformations across projects
\item Avoidance of errors in the transformation process
\item Reduced costs due to decreased complexity
\end{itemize}
As all models used in the presented approach are ontology-based, we follow the principle of Linked Data that allows interlinking and discovery of new data sources in a global data graph \cite{bizer2009linked}. Relying on an established standard for an ontology language, the system also favors sharing and processing of system models used in development. Creating effective Knowledge Management Systems is a key success factor in engineering process improvement \cite{chourabi2008ontologySE}.

% Paper structure
This paper is divided into six sections. This Introduction is followed by a section providing relevant information on the background and related work. Section III describes requirements of a MP transformation system. In section IV the solution approach is presented. Section V describes two use cases. Finally, a conclusion drawn from the experiences made while carrying out this research is given in section VI.

\section{Background and Related Work}
\label{sec:related_work}
This section provides relevant background knowledge. The concepts of MPs and Mission Profile Aware Design (MPAD) are introduced and related work is mentioned. Also, the relation of works from the field of M2M transformation to our approach is described.

\subsection{Mission Profiles}
\label{sec:mission_profiles}
Capturing and specification of environmental conditions and functional loads which components are exposed to throughout their life cycle is done via MPs \cite{byrne2008handbook}. Therefore, they play an important role in the design and development of automotive systems, especially with regard to the propagation of requirements along the development process chain \cite{nirmaier2014mission,abelein2012complexity}. To create a MP various sources are used such as load usage and environmental profiles, requirements and scenarios for stress tests. An initial draft of a format for MPs was developed in the RESCAR2.0 project\footnote{https://www.edacentrum.de/rescar/} and is further described in \cite{nirmaier2014mission}. To improve the collaboration among partners involved in automotive component design the autoSWIFT project\footnote{https://www.edacentrum.de/autoswift/} aimed to prepare a standard for MPs, the Mission Profile Format (MPFO)\footnote{http://www.mpfo.org/}.

%Recently, Hirler et al. presented a theoretical model for reducing stressors in MPs to two single parameters i.e. \emph{effective stress level} and \emph{effective stress time}. Experiments showed that the reliability data obtained fit theoretical predictions within statistical variations \cite{hirler2017evaluation}.

\subsection{Mission Profile Aware Design}
\label{sec:mission_profile_aware_design}
Jerke and Kahng introduced the general concept of MPAD in \cite{jerke2014mission}. They emphasized that MP consideration is still mainly a manual task. This is why it is important to provide tool support to enable automation and ease adoption. Currently, we are only aware of a single framework supporting MPAD, the \emph{Reliability Knowledge Framework} by NXP which was created to connect users, reliability knowledge, data, tools and methods \cite{rongen2014reliabilityFramework}. The rareness of software supporting MPAD might be due to the fact that a MP standard does not exist yet, see section \ref{sec:mission_profiles}.

\subsection{Model-to-Model transformation}
While M2M transformation approaches can be divided into relational/declarative, imperative/operational, graph-based and and hybrid categories \cite{kahani2018survey}
the approach to MP transformation presented in this paper supports multiple M2M transformation approaches and is not limited to one category. This is due to an identified key requirement, which is explained in section \ref{sec:key_requirements}.

\begin{figure*}
  \centering
  \includegraphics[width=.95\linewidth]{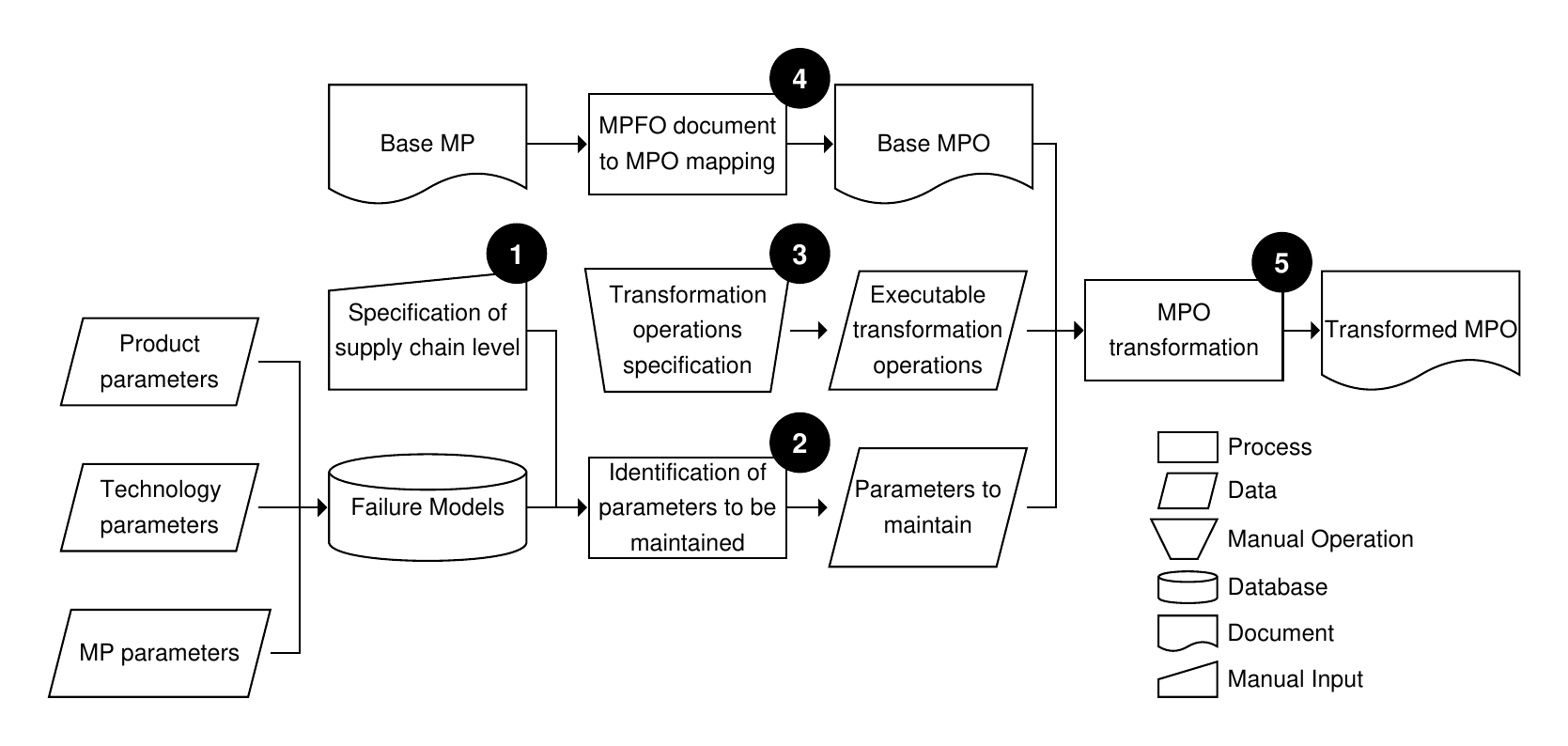}
  \captionsetup{width=\linewidth}
  \caption[bla]{Overview of the approach}
  \label{fig:approach_overview}
\end{figure*}

\section{Requirements}
\label{sec:approach}
In this section we list hypotheses from which we draw vital conclusions in form of identified abstract key requirements. Based on the key requirements, we also define evaluation criteria. This gradual refinement reflects our procedure.

\subsection{Key requirements}
\label{sec:key_requirements}
There are three important hypotheses which we used as a starting point for our considerations.

\subsubsection{Supply chain consideration}
\begin{figure}[H]
  \centering
  \includegraphics[width=\linewidth]{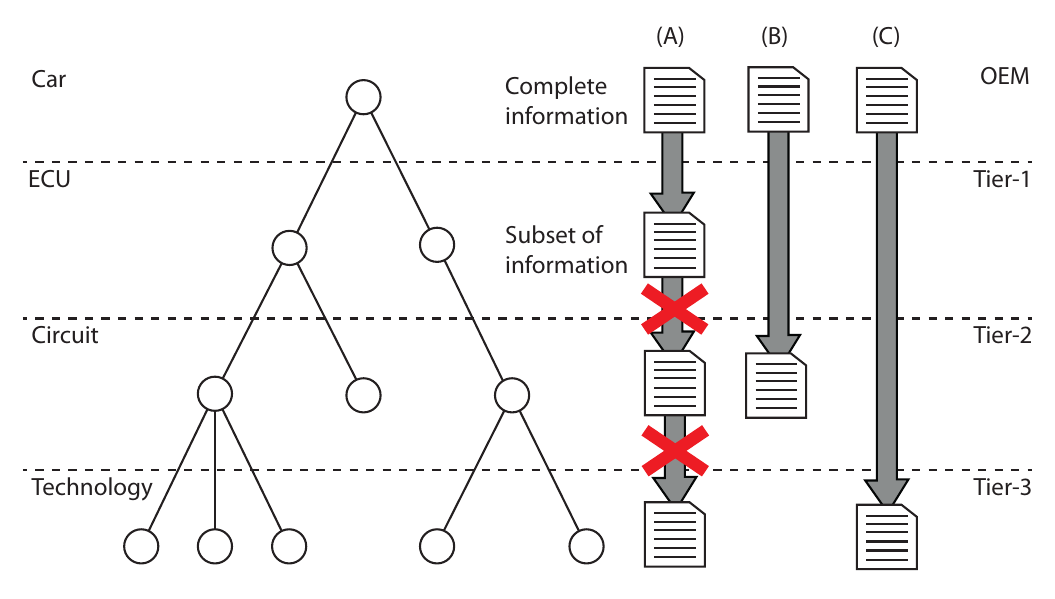}
  \captionsetup{width=\linewidth}
  \caption[bla]{Examplary supply chain and MP propagation. At the OEM level all information is available as MP data. Transformation of the base  MP for the Tier-1 typically results in a subset of MP information. Subsequent transfer of MP data from Tier-1 to Tier-2 is not possible as there might be information missing due to the initial transformation (A). Therefore, a cumbersome process involving transformations of the base MP to each tier's demands is necessary (B), (C)}
  \label{fig:supply_chain_mp_passing_on}
\end{figure}

At various tiers different portions of the original MP data are required. To apply failure models at the technology provider level for example requires maintenance of specific MP parameters at preceding tiers, see Fig. \ref{fig:supply_chain_mp_passing_on}. This can also occur on higher levels. Tools for MP transformation should therefore be aware of the supply chain.

\subsubsection{Generic load specifications}
For providing a general solution to the specification of environmental loads the means to describe the loads have to be versatile. This also applies to the feature set of tools used to process these load specifications. Discovery of previously unknown loads which have an impact on the component in question demands for extensibility of load specification mechanisms and related manipulation functionality of processing tools.

\subsubsection{Varying level of detail}
\label{sec:requirements_level_of_detail}
The level of detail of MPs can vary. Therefore tools for processing MP data need to be scalable. In addition to that, tools for processing MP data should be adaptable in terms of customizable manipulation operations to handle different levels of detail equally well.

\vspace{0.25cm}

We extracted the following abstract key requirements for a MP transformation system from the hypotheses:
\begin{itemize}
\item[\circled{1}] Maintenance mechanism for parameters
\item[\circled{2}] Versatile processing feature set
\item[\circled{3}] Extensible processing feature set
\item[\circled{4}] Customizable manipulation operations
\item[\circled{5}] Scalability
\end{itemize}

\subsection{Evaluation criteria}
The level of abstraction of the former requirements makes it hard to decide what measures are suitable to meet a requirement because it is hard to tell whether it has been fulfilled or not. This also hinders a proper evaluation of the solution later on. To tackle these issues, we defined specific evaluation criteria \fbox{EC} associated with the key requirements which are more precise and specific:
\begin{itemize}[align=left,labelwidth=2cm,labelindent=15pt,leftmargin=62pt,itemsep=0pt]
\item[\fbox{EC.Parameter}] The transformation system contains a mechanism for maintaining parameters required for subsequent tiers
\item[\fbox{EC.Transform}] The transformation system contains at least two transformation approaches which belong to different classes of transformations according to the literature
\item[\fbox{EC.Extension}] The transformation system provides a mechanism to extend it with new transformation approaches
\item[\fbox{EC.Modify}] The transformation behavior of transformation approaches provided by the system is modifiable
\item[\fbox{EC.Detail}] The transformation system should handle different detail levels of MP data equally well
\end{itemize}

\section{Approach}
This section describes our proposal for ontology-based MP transformation. A first section providing an overview of the approach follows a description of the solution concept and a section with a brief summary of the design of a framework which was used to carry out this research. Finally, there is a section describing the architecture of the proposed system.

\subsection{Overview}
Please refer to Fig. \ref{fig:approach_overview} for an overview of the approach. The tier needs to be manually specified \ding{202}. This information is then used in conjunction with known failure models for the identification of parameters to be maintained \ding{203} through the transformation. The manual specification of transformation operations yields executable transformation operations \ding{204}. The MPFO base MP document is mapped to an MP ontology (MPO) using the framework described in section \ref{sec:semantic_mpad_platform} \ding{205}. Finally, the ontology-based transformation of the base MPO is executed \ding{206}.

\subsection{Solution concept}
A fundamental characteristic of our proposed solution is continuous use of a unifying language to express models namely the ontology language OWL. The primary purpose of this is to alleviate difficulties and effort to integrate data, information and knowledge required to fulfill the task of MP transformation and support its implementation.

In order to obtain models expressed in OWL we suggest \emph{lifting} existing models, by which we mean a) translation of the source model into suitable OWL expressions and b) enrichment of the resulting ontology with additional relevant knowledge to support task achievement. Please refer to section \ref{sec:semantic_mpad_platform} for a description of the lifting mechanism we used in this approach to map MPs to ontologies.

The following five paragraphs refer to the previously defined evaluation criteria and outline the measures undertaken to satisfy the key requirements and to achieve the overall objective of our proposal.

\subsubsection{Supply chain awareness}
We achieve consideration of the supply chain through a mechanism for MP parameter maintenance. The idea is to mark parameters as \emph{required} when they are parameters of a failure model that can be applied on a lower tier. This process refers to an ontology-based repository of failure models which is part of the proposed system.

\subsubsection{Multitude of transformation approaches}
To distinguish between transformation approaches we adopt the classification of Czarnecki and Helsen in \cite{czarnecki2003classification}. To fulfill evaluation criterion \fbox{EC.Transform} our proposed system includes a \emph{direct-manipulation} as well as a \emph{relational} M2M transformation approach.
%In addition to that, we include another approach that is called \emph{pattern-based} transformation.

\subsubsection{Extensibility}
Our concept implements extensibility by providing a plug-in mechanism. The concrete design of this mechanism depends on the chosen technology with which the system is implemented. In our case for example, we used a multi-paradigmatic (object-oriented and functional) programming language to implement the transformation system. More transformation approaches can be added to the system by implementing them using predefined interfaces.

\subsubsection{Modifiability}
Our approach considers modifiability of the behavior of transformation approaches in the design of the implemented transformation system. This is primarily achieved by the fact that the specification of a transformation according to the chosen approach already implies a definition of the behavior. In addition, transformation approaches can be parameterized, which has the positive side effect of encouraging reuse.

\subsubsection{Unvarying performance}
To handle different levels of detail in MPs the system relies only on established libraries which are known to perform well. Of course concrete definitions of transformations have a direct impact on performance. Therefore, it is mostly up to the user to keep that in mind while defining transformations.

\subsection{Framework design}
\label{sec:semantic_mpad_platform}
This section describes a framework for the lifting of MP documents to ontologies. We utilized existing tooling of the \emph{Semantic Mission Profile Aware Design Platform} to map MP documents to corresponding ontologies. The platform contains a tool which takes a MP document in MPFO document format as input and generates a corresponding OWL ontology which is serialized in RDF/XML format.

\subsubsection{Architecture and implementation}
\label{sec:framework_architecture_and_implementation}
The platform is implemented as a Play Framework\footnote{https://www.playframework.com/} application using the Scala \cite{odersky2004overview} programming language. This framework is based on the RESTful \cite{fielding2000architectural} architectural design pattern. The idea behind the design of the platform was on the one hand to provide means to create applications supporting MPAD and on the other hand to accelerate development by providing a pragmatic way to actually implement such applications.

\subsubsection{Core functionality}
The primary purpose of the Semantic MPAD Platform is the processing, interlinking, and integration of MPs and requirements. All inputs which are expected to be documents are lifted to ontologies by mapping engines. The platform moreover contains tools to detect and define actions upon changes to these documents. Using the standardized ontology language OWL as output opens up the possibility to easily integrate reasoning mechanisms into the processing of MPs and requirements.

\subsubsection{Interfaces}
Apart from the RESTful API (see section \ref{sec:framework_architecture_and_implementation}) which can be accessed from any application capable of communicating via HTTP, the platform provides a web-based GUI using HTML and JavaScript. This allows interaction with the system from any standard-conforming browser and therefore eases the integration of the system into existing design processes and environments.

\subsection{Architecture}
This section describes the failure model and transformation ontology as well as the architecture of the suggested MP transformation system.

\subsubsection{Failure model and transformation ontology}
\label{sec:failure_model_and_transformation_ontology}
To consider the supply chain and reach evaluation criterion \fbox{EC.Parameter} an ontology addressing the parameter maintenance is introduced. This ontology has the purpose of identifying parameters which need to be maintained through a transformation. The OWL ontology contains SWRL rules for the inference of parameters to maintain. We explain the reasoning utilizing the combination of OWL axioms and SWRL rules by highlighting key axioms and rules of the ontology.

\begin{description}
    \item[$\mathsf{FailureModel}$] The concept of a failure model is modeled by a subclass of the class expressed by axiom \ref{eq:failure_model_superclass}:
    \begin{align}
    \label{eq:failure_model_superclass}
    & \mathsf{\exists hasEquation.xsd\mathord{:}string \sqcap} \notag \\
    & \mathsf{\exists hasMissionProfileParameter.MissionProfileParameter \sqcap} \\
    & \mathsf{\exists ofFailure.Failure} \notag
    \end{align}

    \item[$\mathsf{TechnologicalFailureModel}$] A technological failure model is a specialization and therefore a subclass of a failure model. Moreover, it is also a subclass of the class expressed by axiom \ref{eq:technological_failure_model_superclass}:
    \begin{align}
    \begin{split}
    \label{eq:technological_failure_model_superclass}
    & \mathsf{\exists hasProductParameter.ProductParameter \sqcap} \\
    & \mathsf{\exists hasTechnologyParameter.TechnologyParameter \sqcap} \\
    & \mathsf{\exists ofFailure.TechnologicalFailure \sqcap} \\
    & \mathsf{\mathrel{= 1} hasLayer.Layer}
    \end{split}
    \end{align}

    \item[$\mathsf{MPOTransformation}$] A transformation is modeled as a class which is also a subclass of the class expressed by axiom \ref{eq:mpotransformation_superclass}:
    \begin{align}
    \begin{split}
    \label{eq:mpotransformation_superclass}
    & \mathsf{\geq maintainsParameter.MissionProfileParameter \sqcap} \\
    & \mathsf{\mathrel{= 1} ofSupplyChainLevel.SupplyChainLevel} 
    \end{split}
    \end{align}
\end{description}

The reasoning moreover utilizes SWRL rules to mark parameters which are to be maintained through a transformation. This will be described in the following.

\begin{description}
    \item[$\mathsf{MaintainParametersOEM}$] Rule \ref{eq:maintain_parameter_rule_instance} sets MP parameters which need to be maintained on a $\mathsf{MPOTransformation}$ instance of $\mathsf{OEMLevel}$ when there is a $\mathsf{TechnologicalFailureModel}$ instance which uses these parameters. Depending on the supply chain, rules must be defined analogously for each tier of the supply chain.
    \begin{align}
    \label{eq:maintain_parameter_rule_instance}
    & \mathsf{MPOTransformation(trans) \land} \notag \\
    & \mathsf{ofSupplyChainLevel(trans, OEMLevel) \land} \notag \\
    & \mathsf{TechnologicalFailureModel(fm) \land} \\
    & \mathsf{hasMissionProfileParameter(fm, param)} \notag \\
    & \mathsf{\rightarrow maintainsParameter(trans, param)} \notag
    \end{align}
    
    \item[$\mathsf{MaintainParametersHighLevel}$] We also need to mark all parameters which need to be maintained with regard to high level failure models. Rule \ref{eq:maintain_parameter_high_level_rule} shows that this can be achieved by checking whether a failure model has a root failure residing on the technology level.
    \begin{align}
    \label{eq:maintain_parameter_high_level_rule}
    \begin{split}
    & \mathsf{MPOTransformation(trans) \land} \\
    & \mathsf{ofSupplyChainLevel(trans, OEMLevel) \land} \\
    & \mathsf{HighLevelFailureModel(fm) \land} \\
    & \mathsf{ofFailure(fm, failure) \land} \\
    & \mathsf{hasRootFailure(failure, rootFailure) \land} \\
    & \mathsf{TechnologicalFailure(rootFailure) \land} \\
    & \mathsf{hasMissionProfileParameter(fm, param)} \\
    & \mathsf{\rightarrow maintainsParameter(trans, param)}
    \end{split}
    \end{align}
\end{description}

\begin{figure*}
  \centering
  \includegraphics[width=\linewidth]{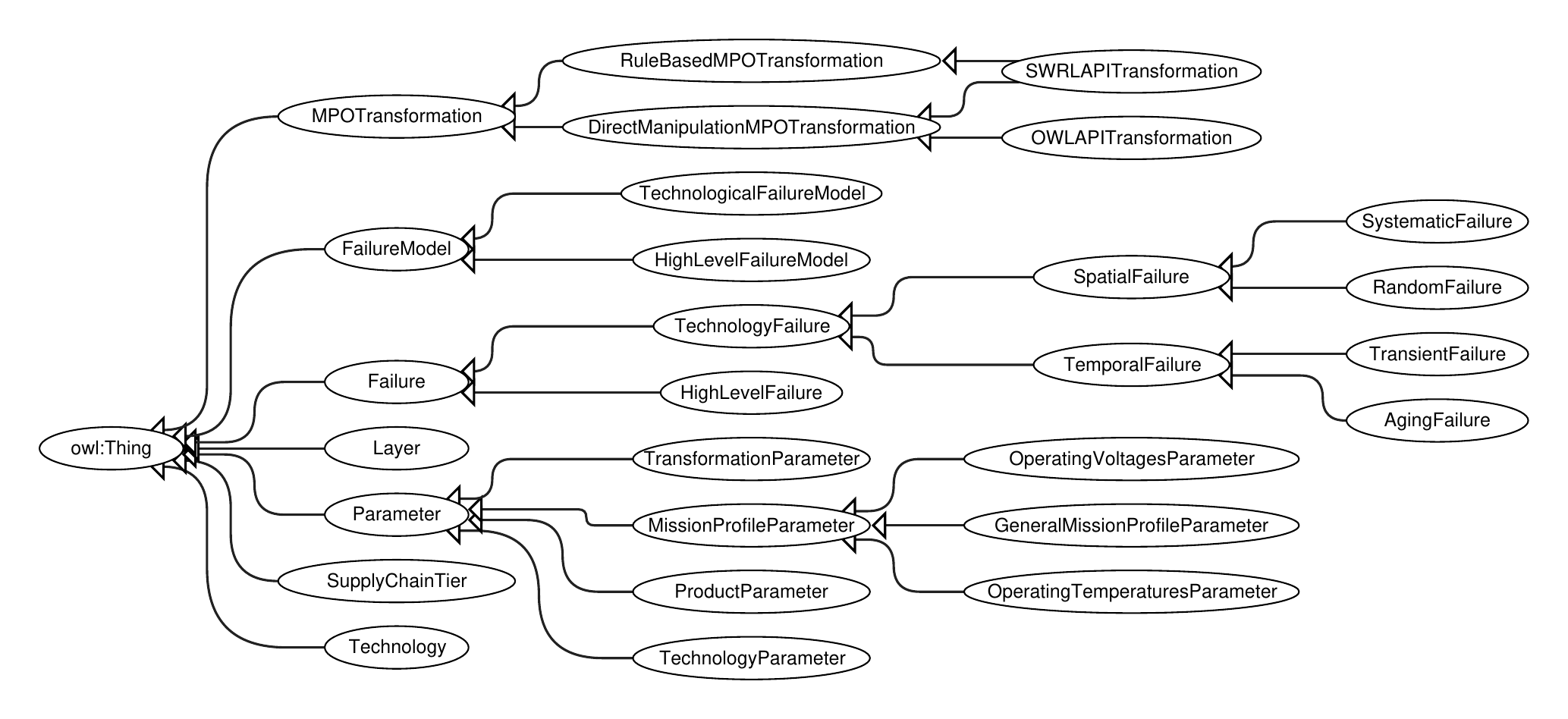}
  \captionsetup{width=\linewidth}
  \caption[bla]{Taxonomy of the failure model and transformation ontology}
  \label{fig:ontology-taxonomy}
\end{figure*}

Furthermore, a taxonomy for failure models was extracted from the categories described by Maricau and Gielen in \cite{maricau2013cmos}. Complementary metal-oxide-semiconductor (CMOS) device failure models for Electromigration (EM), Hot-carrier injection (HCI), Time-dependent dielectric breakdown (TDDB) and Bias temperature instability (BTI) were added as instances of corresponding classes.
Fig. \ref{fig:ontology-taxonomy} shows the taxonomy of the failure model and transformation ontology.

\pagebreak

\subsubsection{System}
\begin{figure}
  \centering
  \includegraphics[width=\linewidth]{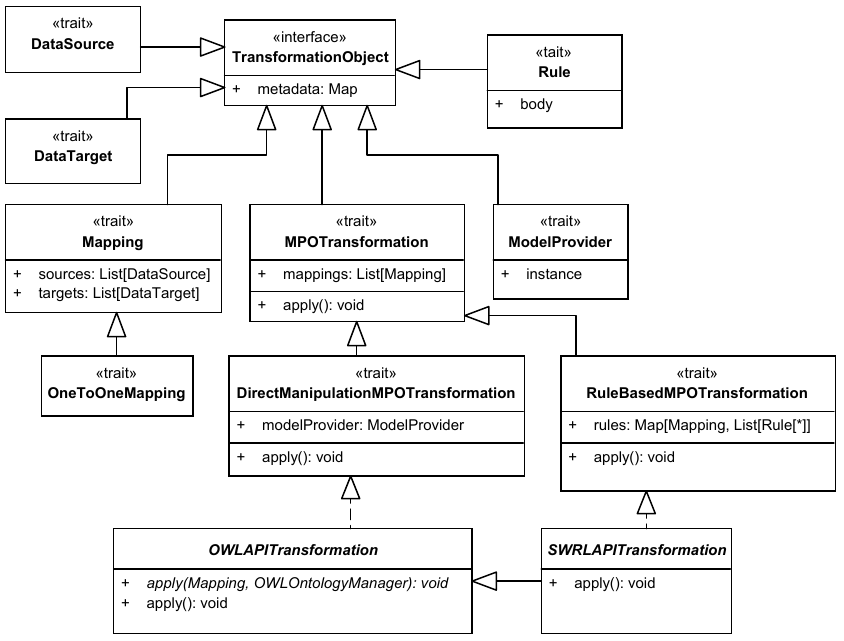}
  \captionsetup{width=\linewidth}
  \caption[bla]{UML diagram of the proposed system}
  \label{fig:system_uml}
\end{figure}

Please refer to Fig. \ref{fig:system_uml} for an overview of the type hierarchy. Please note that we used the Scala programming language which provides so-called \emph{traits} as opposed to interfaces from object-oriented languages such as Java. Scala traits allow multiple inheritance and can have attributes. UML does not have a semantic concept for this type of inheritance. Each object in the system is a \emph{TransformationObject} that provides a \emph{metadata} attribute of arbitrary type, based on key-value access. To meet evaluation criteria \fbox{EC.Transform} and \fbox{EC.Extension}, we introduced a trait for transformations: \emph{MPOTransfromation}. This trait holds \emph{Mappings} which themselves have \emph{sources} and \emph{targets} lists attributes. As an example we have included a \emph{OneToOneMapping} which extends the \emph{Mapping} trait and overrides the sources and targets attributes to provide lists containing only a single element. Designing this way allows the definition of one-to-one, one-to-many, many-to-one and many-to-many mappings. This is required to create the necessary flexibility in MP transformation to cover use cases such as for example merging of several MPs into a single one. Actual data such as file paths, URLs or database addresses needed to use the sources respectively targets can be transported via the metadata.

A \emph{DirectManipulationMPOTransformation} object has a \emph{modelProvider} attribute which is used to access an object providing the model manipulation API as demanded by direct manipulation transformation approaches.
\emph{RuleBasedMPOTransformation} objects contain a \emph{rules} attribute which is basically a mapping of \emph{Mapping} objects to a list of \emph{Rule} objects. This allows specific association of rules to certain mappings. A \emph{Rule} object itself holds a \emph{body} attribute of arbitrary type.
While the trait \emph{OWLAPITransformation} extends \emph{DirectManipulationMPOTransformation}, a \emph{SWRLAPITransformation} extends the latter trait as well as \emph{RuleBasedMPOTransformation}.

To satisfy evaluation criterion \fbox{EC.Modify}, the system lets users define transformation behavior via overriding the \emph{apply} method of transformation traits in implementing classes or specializing traits.

\section{Use cases}
Two typical industrial use cases are described in this section which are concerned with the application of technology failure models utilizing the parameter maintenance mechanism and the derivation of sub-component MPs.
For illustration purposes, both use case descriptions refer to concrete examples.

\subsection{Technology failure model application}
\label{sec:use_case1}
To apply a failure model at the technology level, certain MP parameters need to be maintained in a transformation.
We used a MP document in MPFO version 0.2.1 as an example which contained seven parameters ($t_0$,$t_1$,$t_2$,$v_0$,$v_1$,$v_2$,$failureRate$) and temperature ranges for stress testing a component expressed as eight intervals ranging from -10 °C to 90 °C, see Tab. \ref{tab:intervals}.
\begin{table}
\centering
\begin{tabular}{|l|c|c|}
\hline
\textbf{Interval ID} & \multicolumn{1}{l|}{\textbf{Min value}} & \multicolumn{1}{l|}{\textbf{Max value}} \\ \hline
Interval\_14         & -10.0                                   & 20.0                                    \\ \hline
Interval\_15         & 20.0                                    & 30.0                                    \\ \hline
Interval\_16         & 30.0                                    & 40.0                                    \\ \hline
Interval\_17         & 40.0                                    & 50.0                                    \\ \hline
Interval\_18         & 50.0                                    & 60.0                                    \\ \hline
Interval\_19         & 60.0                                    & 70.0                                    \\ \hline
Interval\_20         & 70.0                                    & 80.0                                    \\ \hline
Interval\_21         & 80.0                                    & 90.0                                    \\ \hline
\end{tabular}
\caption{MP temperature intervals}
\label{tab:intervals}
\end{table}
Prior to actually applying the transformation to the MP, the failure model and transformation ontology is used to find out which parameters need to be maintained. This requires the user to specify the supply chain tier at which the transformation is to be performed. The existence of a knowledge base providing failure models is assumed. In our example, the ontology contains five front-end-of-line (FEOL) and back-end-of-line (BEOL) failure models \emph{FEOL-TDDB}, \emph{FEOL-HCI}, \emph{FEOL-BTI}, \emph{BEOL-TDDB} and \emph{BEOL-EM} with corresponding technology, product and MP parameters defined. This set of MP parameters contained voltage and temperature parameters as well as the failure rate. Utilizing reasoning as described in section \ref{sec:failure_model_and_transformation_ontology}, we found that temperature, voltage, and failure rate parameters needed to be maintained. List. \ref{lst:reasoningExplanation} shows an explanation from the ontology editor Protégé \cite{musen2015protege} for the maintenance of an example temperature parameter.
\begin{figure}[H] % wrap into a figure environment to avoid strange artifacts
\begin{lstlisting}[frame=single,caption=Protégé explanation for an OEM transformation which maintains an example temperature parameter,label=lst:reasoningExplanation,xleftmargin=3.4pt,xrightmargin=3.4pt,morekeywords={SubClassOf,Type},breaklines=true]
writeError hasRootFailure HCI
DirectManipulationMPOTransformation SubClassOf MPOTransformation
example-transform-oem ofSupplyChainLevel oemLevel
example-transform-oem Type SWRLAPITransformation
failureModel-writeError hasMissionProfileParameter exampleMissionProfileParameter_t1
SWRLAPITransformation SubClassOf DirectManipulationMPOTransformation
HCI Type AgingFailure
AgingFailure SubClassOf TemporalFailure
failureModel-wirteError Type HighLevelFailureModel
MPOTransformation(?trans), ofSupplyChainLevel(?trans, oemLevel), HighLevelFailureModel(?fm), ofFailure(?fm, ?failure), hasRootFailure(?failure, ?rootFailure), TechnologicalFailure(?rootFailure), hasMissionProfileParameter(?fm, ?p) -> maintainsParameter(?trans, ?p)
failureModel-writeError ofFailure writeError
TemporalFailure SubClassOf TechnologicalFailure
\end{lstlisting}
\end{figure}

After identifying parameters which need to be maintained, the transformation operations are specified. In our example, we did this by implementing a subclass of the \emph{OWLAPITransformation} trait, please see List. \ref{lst:transformSource} which shows an excerpt of the transformation implementation.
\begin{figure} % wrap into a figure environment to avoid strange artifacts
\begin{lstlisting}[frame=single,caption=Scala source code template for an OWLAPI-based transformation,label=lst:transformSource,xleftmargin=3.4pt,xrightmargin=3.4pt,language=Scala]
class IncreaseTemperatureCycleRange(
    sourcePath: String,
    targetPath: String,
    rangeIncrease: Float) extends OWLAPITransformation {
  override val metadata = Map("description" -> "temperature range increase")
  override val mappings = List(new OneToOneMapping(
    new DataSource { override val metadata = Map("path" -> sourcePath) },
    new DataTarget { override val metadata = Map("path" -> targetPath) })
  )
  override def apply(mapping: Mapping, manager: OWLOntologyManager) = {
    assert(mapping.sources.size == 1 && mapping.targets.size == 1)
    assert(mapping.sources(0).metadata.isDefinedAt("path"))
    assert(mapping.sources(0).metadata("path").isInstanceOf[String])
    assert(mapping.targets(0).metadata.isDefinedAt("path"))
    assert(mapping.targets(0).metadata("path").isInstanceOf[String])
    val source = manager.loadOntologyFromOntologyDocument(
      new File(mapping.sources(0).metadata("path").asInstanceOf[String]))
    val target = manager.createOntology(
      IRI.create("transformed"), Set(source).asJava)
    manager.setOntologyDocumentIRI(target, IRI.create(
      new File(mapping.targets(0).metadata("path").asInstanceOf[String])))
    manager.setOntologyFormat(target, manager.getOntologyFormat(source))
    
    // TRANSFORMATION OPERATIONS
    
    // Apply the changes and save the result
    manager.applyChanges(axiomsToRemove.map(ax =>
      new RemoveAxiom(target, ax)).asJava)
    manager.applyChanges(axiomsToAdd.map(ax =>
      new AddAxiom(target, ax)).asJava)
    manager.saveOntology(target)
  }
}
\end{lstlisting}
\end{figure}
To increase the ranges, we implemented a transformation which decreased the lower and increased the upper limits of the intervals. To execute the transformation, the input MPFO document is mapped to an ontological representation. We used the framework described in section \ref{sec:semantic_mpad_platform} for this task.

\begin{table}[H]
\resizebox{\linewidth}{!}{%
\begin{tabular}{|l|c|c|}
\hline
\textbf{Characteristic} & \textbf{Input ontology} & \textbf{Output ontology} \\ \hline
Axioms & 3874 & 3633 \\ \hline
Logical axioms & 2889 & 2773 \\ \hline
Declaration axioms & 911 & 860 \\ \hline
Class count & 52 & 36 \\ \hline
Object properties & 39 & 31 \\ \hline
Data properties & 30 & 19 \\ \hline
Individuals & 775 & 774 \\ \hline
Object property assertions & 779 & 779 \\ \hline
Data property assertions & 1216 & 1216 \\ \hline
\end{tabular}%
}
\caption{Ontology metrics of input and output MP ontologies}
\label{tab:ontology_metrics}
\end{table}

\pagebreak

Tab. \ref{tab:ontology_metrics} lists metrics of transformation input and output ontologies.
Even though some axioms have been removed in the transformation, the object and data property assertions are preserved. The loss of some axioms is due to intentionally discarded MP information such as operating states and ports specifications.

% - MP data source

%Weather measurement data from Germany's National Meteorological Service, the Deutscher Wetterdienst (DWD) served as dataset. The measurement data was preprocessed using a dedicated MP generator to create a MP document which conforms to the MP format draft, see section \ref{sec:mission_profile_aware_design}. This MP document specifying the environmental conditions to which an arbitrary component was exposed to was then the input for the mapping tool of the Semantic Mission Profile Aware Design Platform described in section \ref{sec:semantic_mpad_platform}. Finally, the resulting ontological representation of the MP was used as a \emph{DataSource} in the system prototype.

\subsection{Sub-component MP derivation}
This use case describes the derivation of a MP for a sub-component from a base MP. As an example, we consider the development of an automotive Electronic Control Unit (ECU). A Tier-1 assembles the ECU module while a Tier-2 provides an Integrated Circuit (IC) component for the ECU which in turn is manufactured using technology provided by a Tier-3. MPs are used in the engineering of the ECU, the IC component and the underlying technology.

%There are various methods to calculate IC junction temperature (REFERENCE).

\begin{figure}[H]
  \centering  \includegraphics[width=\linewidth]{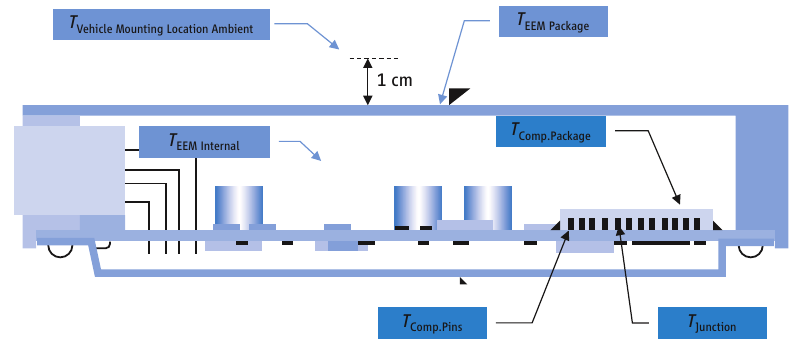}
  \captionsetup{width=\linewidth}
  \caption[bla]{Temperature measurement points of an ECU (here called Electrical/Electronic Module, EEM) defined in the Handbook for Robustness Validation, from \cite{byrne2008handbook}}
  \label{fig:ecu_temperature_measurement_points}
\end{figure}

In our example, the ECU is mounted in the engine compartment. Using a publicly available typical temperature load distribution \cite{hogen2012lifetime} one can define a MP for the ECU. This temperature profile of the inner air temperature of the ECU (referring to Fig. \ref{fig:ecu_temperature_measurement_points}: $T_{\text{EEM Internal}}$) is expressed as a histogram of 12 temperature ranges between -40 °C and 125 °C.
%For a total lifetime of 8000 hours of active operation the temperature range 45 °C to 60 °C for example has the largest share of 2640 hours (33.0 \%).
For providing a MP to the Tier-3 the Tier-1 has to transform the temperature profile MP data.
%The Tier-2 designing the IC is interested in the IC junction temperature (see Fig. \ref{fig:ecu_temperature_measurement_points}: $T_{\text{Junction}}$).

The JEDEC standard \mbox{JESD51-2} defines a methodology for performing thermal resistance measurement for ICs in natural convection. The thermal resistance between device junction and ambient air is determined using equation \ref{eq:junction_temperature_equation_form1}. In its second form in equation \ref{eq:junction_temperature_equation_form2}, given the ambient temperature $T_{\text{A}}$, the total device power dissipation $P$, and the total thermal resistance between device junction and the ambient air $\theta_{\text{JA}}$, one can calculate the junction temperature $T_{\text{J}}$:
\begin{align}
    \label{eq:junction_temperature_equation_form1}
    \theta_{\text{JA}} &= (T_{\text{J}} - T_{\text{A}}) / P & \Leftrightarrow \\
    \label{eq:junction_temperature_equation_form2}
    T_{\text{J}} &= T_{\text{A}} + (P \times \theta_{\text{JA}}) & \Leftrightarrow \\
    \label{eq:junction_temperature_equation_form3}
                    &= \theta_{\text{JC}} + \theta_{\text{CA}}
\end{align}
%where $\theta_{JA}$ is the sum of the thermal resistances $\theta_{JC}$ and $\theta_{CA}$ which 
where $\theta_{\text{JC}}$ is the thermal resistance between junction and case of the device and $\theta_{\text{CA}}$ the thermal resistance between case of the device and ambient air.
Equations \ref{eq:junction_temperature_equation_form2} and \ref{eq:junction_temperature_equation_form3} are also used in the industry for junction temperature calculation\footnote{For example in the tutorial MT-093 from Analog Devices, in the application report SPRA953C from Texas Instruments, the common information document TB379 from Renesas or the introduction to temperature specification document AN4017 from Cypress.}. These equations are  used to calculate a corresponding temperature profile for the Tier-3 based on the ECU MP.

%\pagebreak

Using our proposed system, a MP for the IC sub-component technology can be derived from the base ECU MP while MP parameters required for failure model application are maintained. The actual specification of the transformation operations requires expert knowledge about the specific relationships between the loads of the base component and the sub-component. These relationships are reflected in the transformation. The junction temperature calculation is an example for such relationships.
Parameters which are not identified as required can be stripped off the base MP to protect IP and reduce required storage space. As MPs are required for each sub-component of the ECU the transformation operations specification can be reused thus saving time and avoiding errors.

Suppose one can assume that the failure model and transformation ontology contains (1) information about the supply chain structure and (2) corresponding failure models. In that case an exemplary flow to derive a MP for the IC sub-component under consideration of parameters for failure model application would be:
\begin{enumerate}
    \item The user specifies the supply chain tier at which the transformation is performed in the failure model and transformation ontology for example by using an OWL ontology editor, see List. \ref{lst:individualOEMTransform}.
    
    \begin{figure}[H] % wrap into a figure environment to avoid strange artifacts
    \begin{lstlisting}[frame=single,caption=OWL Individual representing a OEM transformation,label=lst:individualOEMTransform,xleftmargin=3.4pt,xrightmargin=3.4pt]
    Individual: example-transform-oem

    Types: 
        SWRLAPITransformation
    
    Facts:  
     ofSupplyChainLevel  oemLevel
    \end{lstlisting}
    \end{figure}
    
    \item The system identifies parameters which need to be maintained using the failure model and transformation ontology and any OWL reasoner. Parameters to maintain will be shown as objects related to the transformation instance by the \emph{maintainsParameter} property, see List. \ref{lst:individualOEMTransformInferred}.
    
    \begin{figure}[H] % wrap into a figure environment to avoid strange artifacts
    \begin{lstlisting}[frame=single,caption=OWL Individual representing a OEM level transformation in Manchester syntax. Three MP parameters are maintained which is the inferred knowledge,label=lst:individualOEMTransformInferred,xleftmargin=3.4pt,xrightmargin=3.4pt]
    Individual: example-transform-oem
    
        Types: 
            OWLAPITransformation
        
        Facts:  
            ofSupplyChainLevel     oemLevel
            maintainsParameter     MPParam_failureRate
            maintainsParameter     MPParam_v0
            maintainsParameter     MPParam_v1
            maintainsParameter     MPParam_v2
            maintainsParameter     MPParam_t0
            maintainsParameter     MPParam_t1
            maintainsParameter     MPParam_t2
    \end{lstlisting}
    \end{figure}
    
    \item The user specifies the transformation operations via implementing a \emph{MPOTransformation} trait or one of the extended traits. This includes setting data sources and targets of the transformation, see List. \ref{lst:transformSource}. Equations \ref{eq:junction_temperature_equation_form2} and \ref{eq:junction_temperature_equation_form3} are used to actually implement corresponding transformation operations.
    \item The system maps the base MP of the ECU in MPFO format to an OWL ontology. This task is carried out using a component of the platform described in section \ref{sec:semantic_mpad_platform}.
    \item The system finally executes the transformation which results in the IC MP ontology.
\end{enumerate}

\section{Conclusion}
\label{sec:conclusion}
We suggested a transformation system for requirements represented by MPs which is capable of considering the supply chain with respect to high level and technology failure models. The extensible design allows for the integration of multiple transformation approaches which was identified as an important requirement for a MP transformation system.

% Limitations
A limitation of the SWRL rule-based transformation is that SWRL supports monotonic inference only. Because of that SWRL-based transformations can only add more facts to an ontology but neither modify nor remove existing information. This could be mitigated for example by adding pre- and post-processing steps to SWRL transformations. In these extra processing steps facts modification or removal is handled.
Another limiting factor which is related to the current implementation of the component which maps MPs to ontological representations, as described in section \ref{sec:semantic_mpad_platform}, is that the ontology model is constructed in memory. This is problematic for very large MPs which do not occur so frequently in practice.

% Improvements
There are also opportunities to improve the approach. General improvements could consist of approaches for reuse of M2M transformations and verification thereof \cite{kusel2015reuse,rahim2015survey}. An improvement with regard to the set of supported transformation approaches could be integration of pattern-based transformation \cite{wagelaar2010module}.
Further research can also be conducted on direct failure model evaluation using SWRL Built-In extensions.
%General usability of the system could be improved for example by integrating all components through a graphical user interface.

% Takeaway
We considered other technical solutions upfront such as using the Eclipse Modeling Framework (EMF) \cite{steinberg2008emf} instead of ontologies and model transformation languages such as the Epsilon Transformation Language (ETL) \cite{kolovos2008epsilon} or the Atlas Transformation Language (ATL) \cite{jouault2006atl} but decided to keep the system design simple in order to reduce complexity. This  turned out to be a viable way to realize a MP transformation system which confirms previous results of the literature \cite{akehurst2006sitra}.
We used standardized languages and formats which are in the process of being standardized and showed the practical applicability of our approach.

\section{Acknowledgement}
This paper is partially funded by the BMBF within the project SAFE4I (grant number 01\textbar{}S17032C), and by the BMWi within the project InsightProducts (grant number 228EN/2).


\balance

\bibliographystyle{IEEEtran}
\bibliography{main}

\end{document}